# ATTENTION DOES NOT GUARANTEE BEST PERFORMANCE IN SPEECH ENHANCEMENT

Zhongshu Hou[1,2,3], Qinwen Hu[1,2,3], Kai Chen[1,2,3], Jing Lu[1,2,3]

[1] Key Laboratory of Modern Acoustics, Nanjing University, Nanjing 210093, China
[2] NJU-Horizon Intelligent Audio Lab, Horizon Robotics, Beijing 100094, China
[3] Nanjing Institute of Advanced Artificial Intelligence, Nanjing 210014, China

{zhongshu.hou, qinwen.hu}@smail.nju.edu.cn, {chenkai, lujing}@nju.edu.cn

## ABSTRACT

Attention mechanism has been widely utilized in speech enhancement (SE) because theoretically it can effectively model the long-term inherent connection of signal both in time domain and spectrum domain. However, the generally used global attention mechanism might not be the best choice since the adjacent information naturally imposes more influence than the far-apart information in speech enhancement. In this paper, we validate this conjecture by replacing attention with RNN in two typical state-of-the-art (SOTA) models, multi-scale temporal frequency convolutional network (MTFAA) with axial attention and conformer-based metric-GAN network (CMGAN).

## 1. INTRODUCTION

MTFAA, which ranks 1st in DNS4 [1], and CMGAN, which achieves the best performance [2] on the public VoiceBank+DEMAND dataset [3], are regarded as two SOTA SE models. They both utilize attention modules. In this paper, we replace the attention modules in MTFAA both in time domain and frequency domain with RNN, and replace the attention modules in CMGAN in frequency domain with RNN, to verify the efficacy of attention by comparing the performance of the systems before and after the substitution.

## 2. EXPERIMENTS

### 2.1. Datasets

We evaluate the system performance on the DNS4 [4] and DCASE [5] datasets where the clean clips are selected from the DNS4 speech dataset and noise recordings are from DCASE dataset. Room impulse responses are convolved with clean speech to generate simulated reverberant speech and the early reverberant speech is preserved as training target. Reverberant utterances are mixed with noise clips at SNRs ranging from -15 dB to 0 dB in the training stage and -15dB to 15dB in the testing stage. All utterances are sampled at 16 kHz in our experiments.

### 2.2. Parameter setup and training strategy

The window size and hop length of short time Fourier transformation (STFT) are 32ms and 8ms, respectively. The discrete Fourier transformation length is 512 and the Hanning widow is used for overlap-add waveform reconstruction. The axial attention modules are replaced by dual path recurrent neural network (DPRNN) [6] in MTFAA (indicated as MTFAA-DPRNN) and the attention modules in CMGAN in frequency domain are substituted with conventional RNN (indicated as CMGAN-FRNN). Other parameters of MTFAA and CMGAN can be found in [1] and [2], respectively.

The batch size in our training is 8 and the temporal length of input audio is 8 s. Warmup strategy [7] is critical in training self-attention based model, where the learning rate $\alpha$ is updated with the rule: $\alpha = \frac{1}{\sqrt{C}} \times \min\left(\frac{1}{\sqrt{\varphi}}, \frac{\varphi}{\sqrt{\Psi^3}}\right)$, with $C = 64$, warmup steps $\Psi = 8000$ and $\varphi$ denoting the training step. We train the model by the warmup-based Adam optimizer with $\beta_1 = 0.9$, $\beta_2 = 0.98$, and $\epsilon = 10^{-9}$. The compression parameter $\gamma$ is $\frac{1}{3}$.

### 2.3. Evaluation metrics

We choose a set of commonly used speech metrics to evaluate the comprehensive performance, i.e., perceptual evaluation of speech quality (PESQ) [8], short-time objective intelligibility (STOI) [9], scale-invariant signal distortion ratio (SiSDR) and composite mean opinion score (MOS) based metrics [10] including MOS prediction of the signal distortion (CSIG), MOS prediction of the intrusiveness of background noise (CBAK) and MOS prediction of the overall effect (COVL). Higher values indicate better performance for all metrics.

### 2.4. Experimental results

Table 1. Experiment results

| Metrics | PESQ | | | | STOI | | | | SDR(dB) | | | |
|---|---|---|---|---|---|---|---|---|---|---|---|---|
| SNR(dB) | -15~-5 | -5~5 | 5~15 | Avg. | -15~-5 | -5~5 | 5~15 | Avg. | -15~-5 | -5~5 | 5~15 | Avg. |
| Noisy | 1.122 | 1.052 | 1.142 | 1.105 | 0.432 | 0.628 | 0.793 | 0.617 | -12.849 | -2.540 | 5.131 | -3.419 |
| MTFAA | 1.215 | 1.657 | 2.163 | 1.678 | 0.591 | 0.803 | 0.891 | 0.762 | 0.131 | 7.093 | 10.263 | 5.829 |
| MTFAA-DPRNN | 1.291 | 1.831 | 2.396 | 1.839 | **0.627** | 0.831 | 0.910 | 0.789 | **1.672** | 8.380 | **11.699** | **7.250** |
| CMGAN | 1.279 | 1.797 | 2.375 | 1.817 | 0.617 | 0.828 | 0.907 | 0.784 | 1.340 | 8.323 | 10.842 | 6.835 |
| CMGAN-FRNN | **1.321** | **1.876** | **2.460** | **1.886** | 0.625 | **0.836** | **0.913** | **0.791** | 1.403 | **8.394** | 11.309 | 7.035 |

| Metrics | CSIG | | | | CBAK | | | | COVL | | | |
|---|---|---|---|---|---|---|---|---|---|---|---|---|
| SNR(dB) | -15~-5 | -5~5 | 5~15 | Avg. | -15~-5 | -5~5 | 5~15 | Avg. | -15~-5 | -5~5 | 5~15 | Avg. |
| Noisy | 1.098 | 1.333 | 1.926 | 1.452 | 1.036 | 1.240 | 1.724 | 1.333 | 1.036 | 1.113 | 1.447 | 1.199 |
| MTFAA | 2.058 | 2.821 | 3.376 | 2.752 | 1.801 | 2.240 | 2.593 | 2.211 | 1.540 | 2.184 | 2.738 | 2.154 |
| MTFAA-DPRNN | **2.252** | **3.075** | 3.650 | 2.992 | 1.890 | 2.400 | 2.775 | 2.355 | **1.680** | 2.411 | 3.001 | 2.364 |
| CMGAN | 2.151 | 3.029 | 3.678 | 2.952 | 1.890 | 2.518 | 2.941 | 2.450 | 1.624 | 2.370 | 3.009 | 2.334 |
| CMGAN-FRNN | 2.218 | 3.056 | **3.708** | **2.994** | **1.909** | **2.557** | **2.987** | **2.484** | 1.679 | **2.425** | **3.067** | **2.390** |

It can be seen from Table 1 that both MTFAA and CMGAN achieve better performance on all metrics after replacing attention with RNN.

## 4. CONCLUSIONS

In this paper, we replace the attention modules in two SOTA SE models with RNN and validate that attention doe not guarantee the best performance.